# An Exploration of Symmetries in the Friedmann Equation

Michael Ibison[a]

[a] *Institute for Advanced Studies at Austin, 11855 Research Boulevard, Austin, TX 78759, USA*

**Abstract.** The Friedmann Equation for a conformal scale factor $a(t)$ is observed to be invariant under a Mobius transformation. Using that freedom, a synthetic scale factor $z(t)$ is defined that obeys a modified Friedman equation invariant under the replacement $z(t) \to \pm 1/\, z(t)$. If this is taken this to be the more fundamental form then the traditional Friedmann equation can be shown to be missing a term due to a species with equation of state $w = -2/3$. We investigate in detail one particular cosmology in which it is possible to specify the contribution from this new species.

We suggest a means of avoiding a potentially redundant copy of the development of the universe the above implies through a cosmological spacetime manifold that is a Mobius band closed in time and non-orientable in space. Though it is closed, a Dirac field in such a spacetime may still possess a global arrow of time by virtue of the twist of the Mobius band.

## INTRODUCTION

In this paper we investigate the symmetries intrinsic to the Friedmann equation and explore possible interpretations. One of these is that the manifold of spacetime in the large is closed in time, with the future conformal singularity and the Big Bang occupying the same location. The Conformal Cyclic Cosmology recently promoted by Penrose [1] expresses a similar notion, though for him these two events are physically similar but not identical; though the Penrose Cosmology is cyclic its manifold is not closed in time.

## COSMOLOGICAL METRICS

The Robertson-Walker (RW) metrics indexed by $K \in [-1,0,1]$ can be expressed as line elements

$$ds^2 = d\tau^2 - a^2(\tau)\left(dr^2 + \left(\sin(kr)/k\right)^2 d\Omega^2\right) \tag{1}$$

where respectively, $k$ is real, zero, or imaginary. Here we will prefer to express these in the form

$$ds^2 = a^2(t)\left(dt^2 - dr^2 - \left(\sin(kr)/k\right)^2 d\Omega^2\right) \tag{2}$$

where $a(t)dt = d\tau$. We refer to $a(t)$ as it appears in (2) as the RW scale factor. All three line elements can be expressed in conformal form [2]

$$ds^2 = A^2(t,r;k)\left(dt^2 - dr^2 - r^2 d\Omega^2\right) \tag{3}$$

where $t$, $r$ in (2) and (3) are different coordinates. In flat space ($k$ = 0) however, $a(t) = A(t,r;0)$. The subject of the Friedmann equation is the RW scale factor (and therefore the conformal scale factor if $k$ = 0). The relationship between the RW scale factor and the conformal scale factor is given in the Appendix.

## MATTER AND RADIATION IN A CONFORMALLY-EXPRESSED SPACETIME

Electromagnetism in curved spacetime expressed in conformal coordinates can be cast – with a particular choice of gauge – as completely independent of the conformal factor, though that invariance is lost in a covariant gauge [3]. *Electrodynamics*, however, always depends on the conformal factor because in both the classical and Dirac theories expressed in conformal coordinates the rest mass behaves as $m_0 \to m_0 A_K(t,r)$ where $A_K(t,r)$ is the conformal scale factor for the RW geometry indexed by K [4]:

Though it corresponds to the infinite future as measured by the Robertson-Walker time variable, the conformal singularity occurs at a finite conformal time, about 15.2 GYr from the present era [3]. Due to their behavior in conformal coordinates, conformal time can be said to be the natural measure of time for EM fields.[1] We will see below that the evolution of the scale factor as described by the Friedmann equation has a redundancy such that the post-singularity evolution may be interpreted as an image of the pre-singularity evolution. If this interpretation is enforced then both matter and EM fields are subjected to a physical boundary condition at the singularity [3,4].

## FRIEDMANN PARAMETERS

For the line element (2), and assuming a homogeneous fluid, GR gives that [5]

[1] Conformal time is the time kept by a clock whose tick is the bounce of a light pulse confined to a pair of parallel mirrors moving with - and therefore separating with - the Hubble flow.

$$\frac{3}{8\pi G a^4}\left(\left(\frac{da}{dt}\right)^2 + k^2 a^2\right) = \frac{\rho_r}{a^4} + \frac{\rho_m}{a^3} + \rho_v \tag{4}$$

where $\rho_i$ is the (fixed) energy density of the $i^{th}$ species at the time that $a$ = 1. Subscripts $r$, $m$, $v$ stand for radiation, matter and vacuum respectively. Here 'matter' includes both baryonic and cold dark matter (CDM). Let us make the definitions

$$H := \left.\frac{da}{dt}\right|_{a=1}, \quad \Omega_c := -k^2 / H^2, \quad \Omega_i := \frac{8\pi G}{3H^2}\rho_i; \quad i \in [r, m, 3, v] \tag{5}$$

so that (4) becomes

$$\frac{1}{H^2}\left(\frac{da}{dt}\right)^2 = \Omega_r + \Omega_m a + \Omega_c a^2 + \Omega_3 a^3 + \Omega_v a^4 \tag{6}$$

We have taken the opportunity to introduce an $\Omega_3$ which is zero if (4) is correct, but is required in later discussions. It follows from these definitions that

$$\Omega_r + \Omega_m + \Omega_c + \Omega_3 + \Omega_v = 1\,. \tag{7}$$

Hence defining an $\Omega$ that does not include the curvature term

$$\Omega := \Omega_r + \Omega_m + \Omega_3 + \Omega_v \tag{8}$$

one has $\Omega - 1 = \Omega_c$, which is the critical observational parameter to determine spatial curvature. Estimates of these parameters obtained from the Particle Data Group [6] are

$$\begin{aligned} h &= 0.72 \pm 0.03 \\ \Omega_b h^2 &= 0.0227 \pm 0.0006 \Rightarrow \Omega_b = 0.044 \pm 0.004 \\ \Omega_m h^2 &= 0.133 \pm 0.006 \Rightarrow \Omega_m = 0.26 \pm 0.02 \\ \Omega_r h^2 &= 2.47 \times 10^{-5} \Rightarrow \Omega_r = 4.8 \times 10^{-5} \pm 4 \times 10^{-6} \\ \Omega_v &= 0.74 \pm 0.03 \end{aligned} \tag{9}$$

where the radiation component is presumed due predominantly to the Cosmic Microwave Background. Subscript $b$ denotes the baryonic (visible) component of matter. Another parameter of interest here is the (time-varying) deceleration parameter $q$, which in RW coordinates is defined to be

$$q := -\frac{a_{\tau\tau} a}{a_\tau^{\;2}}; \quad d\tau = a\,dt \tag{10}$$

In conformal coordinates therefore it is

$$q := -\frac{\frac{d}{dt}\left(\frac{1}{a}\frac{d}{dt}a\right)}{\left(\frac{1}{a}\frac{d}{dt}a\right)^2} = -\frac{\ddot{a}/a-(\dot{a}/a)^2}{(\dot{a}/a)^2} = 1-\frac{a\ddot{a}}{\dot{a}^2} \tag{11}$$

# THE CONFORMAL SINGULARITY

The scale factor solving the Friedmann equation passes through $a=0$ (the Big Bang) at some earlier and finite conformal time (with respect to the present). This is not a mathematical singularity – the scale factor and its derivative are well-defined and finite there. Additionally, there is a final singularity $a=+\infty$ at some finite time approached from below, provided the scale factor grows sufficiently that the vacuum term eventually dominates. In the limit, the behavior is independent of the presence or absence of the curvature term. With reference to (A4) and (A9) the conformal scale factor is likewise singular when $a(t)$ is singular. And with reference to (A1) the conformal time *T* (*t* in (3)) is related to the Robertson-Walker time *t* in (2) by

$$K(T-T_0) = \tan\left(\frac{k}{2}(t-t_0+r)\right)+\tan\left(\frac{k}{2}(t-t_0-r)\right). \tag{A12}$$

If we choose the offset $t_0$ to be the time when $a(t)$ is singular, i.e. so that $a(0)$ is singular, then $T=T_0$ at the singularity, which is also finite, and is independent of R. Thereafter the discussion of the curved-space and flat-space Cosmologies can proceed in parallel. For observed values of the $\Omega_i$ (subject to the presumption $\Omega_3$ = 0), the scale factor evolves towards a de-Sitter asymptote

$$\frac{1}{H^2}\left(\frac{da}{dt}\right)^2 \to \Omega_v a^4 \Rightarrow a \to \frac{1}{1-t/t_{cs}} \tag{13}$$

where $t_{cs}$ is the (conformal) time of the conformal singularity. To be specific we will put the big bang at time $t_{bb}$: $a(t_{bb})=0$.

Suppose the Friedmann equation is written

$$\frac{1}{H^2}\left(\frac{da}{dt}\right)^2 = p(a) \tag{14}$$

where $p(a)$ is presumed expressible as a finite Taylor Laurent Series about $a=0$:

$$p(a) = \sum_{j=m}^{n} c_j a^j \tag{15}$$

Then the time from the present era to the conformal singularity is

$$t_{cs} - t_0 = \int_1^\infty da \frac{1}{H\sqrt{p(a)}} \tag{16}$$

is the time from the present era to the conformal singularity, if one exists. A necessary condition for this time to be finite is that $n \geq 3$. For the singularity to be accessible from the present time requires that $p(a) = 0$ remains positive in the interval $a \in [1, \infty]$ for which it is sufficient but not necessary that the $c_j$ are all non-negative and either or both of $c_3, c_4$ are non-zero. Both conditions are met of course by the traditional values incorporated into (6) when $c_3 = \Omega_3 = 0,\ c_4 = \Omega_v > 0$. Accepted values for the **Ω** give $t_{cs} - t_0 = 15.2$ GYr. These conditions are not met when the metric is written in RW time (with line element (1)) because in that case the vacuum term corresponds to $n = 2$, $c_2 = \Omega_v$. Correspondingly, the future singularity is infinitely distant in time.

The conformal time since the Big Bang is

$$t_0 - t_{BB} = \int_0^1 da \frac{1}{H\sqrt{p(a)}} \tag{17}$$

A sufficient condition that this is finite is that $m \geq 0$ and that the $c_j$ are all non-negative (and at least one is non-zero). This condition is also met in practice, so the conformal time to the Big Bang is finite, and computed to be 47.19 GYr using accepted values for the Ω.

The total conformal duration, i.e. the time from the Big Bang to the conformal singularity, is finite and equal to about 62.4 GYr. Inflation is absent from these considerations. In practice inflation turns out not to change the conclusion that the time since the Big Bang is finite, though it does of course increase that interval.

## SYMMETRIES OF THE FRIEDMANN EQUATION

### Form-Invariance Under Reciprocity

We draw attention at first to the fact that Eq. (4) retains its form under the replacement

$$a(t) \rightarrow \pm 1 / a(t) \tag{18}$$

which we will call form invariance. The invariance is true only of the conformal form. In RW coordinates with line element (1) the Friedmann equation is

$$\frac{3}{8\pi G}\left(\left(\frac{da}{d\tau}\right)^2 + k^2\right) = \frac{\rho_r}{a^2} + \frac{\rho_m}{a} + a^2 \rho_v \tag{19}$$

which (18) does not leave unchanged in form. In the following we will be interested in an interpretation of the invariance of (6) with respect to (18) as indicative of a more than usual equitable treatment of contra-variant and covariant coordinates. Defining $b(t)=1/a(t)$, then $b(t)$ is the conformal scale factor for the covariant coordinates:

$$ds^2 = g_{\mu\nu}dx^\mu dx^\nu = a^2(t)\eta_{\mu\nu} = g^{\mu\nu}dx_\mu dx_\nu = b^2(t)\eta^{\mu\nu} \tag{20}$$

What significance then should we attach to the fact that the Friedmann equation is form invariant under $a \leftrightarrow \pm b$, an exchange that swaps the Big Bang with the conformal singularity? Does this suggest a relationship between the two? Note though that the effects of $a \leftrightarrow \pm 1/a$ on (6) are

$$\frac{1}{H^2}\left(\frac{da}{dt}\right)^2 = \Omega_r a^4 + \Omega_m a^3 + \Omega_c a^2 + \Omega_3 a + \Omega_v \tag{21}$$

which, given the observed values of Ω, is not the same as (6), and therefore any such relationship, if it exists is not at the level of the scale factor alone.

## Mobius Transformation

### *Form Invariance*

A generalization of (18) is the Mobius Transformation

$$a(t) = \frac{\alpha + \beta z(t)}{\gamma + \delta z(t)} \tag{22}$$

Note that there are only 3 effective constant degrees of freedom in (22). A convenient choice is to set $\delta\alpha - \beta\gamma = 1$. With this normalization putting (22) into (6) gives

$$\begin{aligned}\frac{1}{H^2}\left(\frac{dz}{dt}\right)^2 &= \Omega_r(\gamma+\delta z)^4 + \Omega_m(\gamma+\delta z)^3(\alpha+\beta z) + \Omega_c(\gamma+\delta z)^2(\alpha+\beta z)^2 \\ &+ \Omega_3(\gamma+\delta z)(\alpha+\beta z)^3 + \Omega_v(\alpha+\beta z)^4\end{aligned} \tag{23}$$

which we will write as

$$\frac{1}{H^2}\left(\frac{dz}{dt}\right)^2 = \sum_{i=0}^{4} \Gamma_i z^i \tag{24}$$

where

$$\Gamma_i = \Gamma_i(\alpha,\beta,\gamma,\delta,\Omega_r,\Omega_m,\Omega_c,\Omega_3,\Omega_v)\,\forall i \in [0,4]\,. \tag{25}$$

Eq. (24) demonstrates that the conformal Friedmann equation is form-invariant under an arbitrary Mobius transformation. Note that by virtue of (22) each term on the right hand side in (24) is due to a combination of contributions from different species (matter, radiation, …), each the coefficient of a single stress-energy tensor and equation of state.

## *Exact Reciprocal Invariance: 'Self-Similarity'*

Here we exploit the generality of the invariance under the transformation (22) and choose the constants so that the form invariance under $a \leftrightarrow \pm 1/a$ is promoted to an exact invariance, here as $z \leftrightarrow \sigma_z / z$.[2] With reference to (24) this requires

$$\Gamma_0 = \Gamma_4, \quad \Gamma_1 = \sigma_\Gamma \Gamma_3, \quad \sigma_\Gamma = \pm 1 \tag{26}$$

which gives

$$\frac{1}{H^2}\left(\frac{dz}{dt}\right)^2 = \Gamma_0\left(1+z^4\right) + \Gamma_1\left(z+\sigma_\Gamma z^3\right) + \Gamma_2 z^2 \tag{27}$$

Now setting $z = \sigma_z / \xi$ in that equation gives

$$\frac{1}{H^2}\left(\frac{d\xi}{dt}\right)^2 = \Gamma_0\left(1+\xi^4\right) + \sigma_z \sigma_\Gamma \Gamma_1\left(\xi+\sigma_\Gamma \xi^3\right) + \Gamma_2 \xi^2 \tag{28}$$

Hence if $\sigma_\Gamma = \sigma_z = \sigma$ then (27) is invariant under $z \leftrightarrow \sigma_z / z$. We will refer to this invariance as 'self-similarity' (under a reciprocal transformation).

From counting the constraints (2) and the degrees of freedom (3), self-similarity can be achieved on the Mobius-transformed Friedmann equation by choice of the constants $\alpha,\beta,\gamma,\delta$ alone, without a constraint on the energy densities $\Omega_i$. It is concluded that the Friedmann equation is self-similar following an appropriately chosen Mobius transformation.

## *Interpretation*

Given the above, one can take the view that the Friedmann equation is not *specifically* a description of the evolution of the scale factor for the contravariant coordinates, but is instead a relatively democratic function the two scale factors associated with the contravariant and covariant coordinates. Inverting (22) one has

$$z(t) = -\frac{\alpha - \gamma a(t)}{\beta - \delta a(t)} = -\frac{\gamma - \alpha b(t)}{\delta - \beta b(t)} \tag{29}$$

which emphasizes the democratic roles of *a* and *b*. Further, we can take the position that the Friedmann equation is fundamentally the self-similar equation (27) in $z(t)$, and that the traditional rendering in terms of $a(t)$ is the projection (22), using particular values for $\alpha, \beta, \gamma, \delta$.

---

[2] This relation is also adopted by Penrose, though his hypothesis presumes the creation of a new cycle rather than a manifold closed in time as presumed here. However that distinction makes no difference to the argument that follows.

## *Relation between the Roots*

The roots of the quartic in *z* on the right hand side of (27) are simply related given that the equation is self-similar. Investigating just the case $\sigma$ = -1 for now, write (27) as

$$\frac{1}{H^2}\left(\frac{dz}{dt}\right)^2 = \Gamma_0\left(1 + z^4 + 2k\left(z - z^3\right) + \left(\kappa^2 - \lambda^2 - 2\right)z^2\right) \tag{30}$$

where

$$2\kappa = \Gamma_1 / \Gamma_0, \quad \lambda = \sqrt{\kappa^2 - 2 - \Gamma_2 / \Gamma_0} \tag{31}$$

and define

$$u = (\kappa + \lambda)/2, \quad v = (\kappa - \lambda)/2 \tag{32}$$

then (30) is

$$\frac{1}{H^2}\left(\frac{dz}{dt}\right)^2 = \Gamma_0\left(z + u + \sqrt{1+u^2}\right)\left(z - u + \sqrt{1+u^2}\right)\left(z + v + \sqrt{1+v^2}\right)\left(z - v + \sqrt{1+v^2}\right) \tag{33}$$

The roots are now in reciprocal pairs:

$$\frac{1}{u + \sqrt{1+u^2}} = -\left(u - \sqrt{1+u^2}\right) \tag{34}$$

Hence

$$\frac{1}{H^2}\left(\frac{dz}{dt}\right)^2 = \Gamma_0\left(z + \zeta(u)\right)\left(z - 1/\zeta(u)\right)\left(z + \zeta(v)\right)\left(z - 1/\zeta(v)\right) \tag{35}$$

where

$$\zeta(x) = x + \sqrt{1+x^2} \tag{36}$$

# THE $\Omega_3$ CONTRIBUTION

## Justification

As we have said, there are 3 degrees effective of freedom in (22) and (24), so the self-similarity condition as $z \to \pm 1/z$ that $\Gamma_0 = \Gamma_4$, $\Gamma_1 = \pm\Gamma_3$ can be satisfied even if $\Omega_3 = 0$. (The $\Gamma_2$ curvature term is automatically self-similar.) If however we adopt the position suggested above that the Friedmann equation is fundamentally self-similar, then the projection using (22) onto the traditional form will not, in general result in $\Omega_3 = 0$. From that standpoint the tradition of $\Omega_3 = 0$ in (21) must be questioned; in general one then expects to find evidence for a term in the traditional Friedmann equation proportional to $a^3$.

## Equation of State

In (6) the stress-energy for each contribution is diagonal and homogeneous:

$$\{T^a{}_b\} = diag(\rho,-p,-p,-p) = \rho diag(1,-w,-w,-w); \quad w = p/\rho . \tag{37}$$

According to the above argument the Friedmann equation is fundamentally self-similar, then $T$ now generates a term in that equation for each of $\Omega_i a^i(t)$ because it does so for each of $\Gamma_i z^i(t)$ in (27) where $i = trace(T)/\rho$. For vacuum and radiation then

$$T_v = \rho_v diag(1,1,1,1); \quad w_v = -1, \qquad T_r = \rho_r diag(1,-1/3,-1/3,-1/3); \quad w_r = 1/3 \tag{38}$$

whilst for the hypothetical $T_3$ component:

$$T_3 = \rho_3 diag\left(1,\frac{2}{3},\frac{2}{3},\frac{2}{3}\right); \quad w_3 = -2/3 \tag{39}$$

## Prediction

It would be interesting to re-examine the CMB and SN data taking (6) as the proper Friedmann equation, allowing for all 5 terms, whilst allowing for the influence of a new species of stress-energy having equation of state (39). An ideal outcome would be that $T_3$ replace the role of dark matter in structure formation. (If MOND can account for anomalous galaxy rotations [7,8] then perhaps dark matter can be excised altogether.) Even if this turns out not to be the case, nonetheless, the conjecture above regarding self-similarity predicts a non-vanishing contribution from the $\Omega_3$ species.

## PERIODICITY

The Friedmann equation (6) and more generally the form (14) is invariant with respect to time translation. Therefore if $a(t)$ is a solution, then so is $a(t+\tau)$ for any fixed $\tau$. As we have said, in practice the time interval between the times at which the scale factor is zero and infinite is finite. The qualitative behavior of $z(t)$ solving the self-similar equation (27) may be the same or different, depending on the coefficients $\Gamma_0, \Gamma_1, \Gamma_2$, and on $\sigma_\Gamma$. We will consider here the simplest case that all of these coefficients are such that $\Gamma_2 > 0$ and

$$\Gamma_0\left(1+z^4\right)+\Gamma_1\left(z+\sigma_\Gamma z^3\right)+\Gamma_2 z^2 \geq 0 \quad \forall z \in [0,\infty] \tag{40}$$

In that case $z(t)$, like $a(t)$, will have a solution that evolves from $z=0$ to $z=+\infty$ in a finite time, where the asymptotic behavior is of the de Sitter kind, $z \sim 1/(t_{cs}-t)$ for some constant $t_{cs}$.

Passing through the singularity $z(t)$ becomes negative infinite. To determine the subsequent post-singularity behavior it is useful then to exploit the self-similarity by letting $z=\sigma_\Gamma/\xi$. This converts (27) back into itself whilst mapping the immediately post-singularity value of $z(t_{cs+})$ onto a Big Bang in $\xi(t)$ at that time. Subsequently $\xi(t)$ must therefore follow the same development as did $z(t)$, though delayed by the duration $\tau := t_{cs}-t_{BB}$. Then

$$\xi(t)=z(t-\tau)=\sigma_\Gamma / z(t) \tag{41}$$

which also implies

$$z(t)=z(t-2\tau) \tag{42}$$

Hence the solutions to the self-similar equation are periodic, with period of twice the duration. It follows from (22) that if $z(t)$ is periodic then so is the scale factor $a(t)$.[3]

## The 'No Copy' Assumption

Adjacent intervals of duration $\tau=t_{cs}-t_{BB}$ are half-cycles of the greater cycle. Let

$$z(t)=\sqrt{\sigma_\Gamma}\exp\left[e^{i\pi t/\tau} f(t)\right] \tag{43}$$

[3] In [4] it was assumed that the energy density changes sign with the scale factor, in which case the mass term in (6) would generalize to $\left|\Omega_m a(t)\right|$. By contrast in the above derivation of periodicity the sign of $\Omega_m$ is assumed to be independent of the scale factor.

then (41) requires

$$f(t) = f(t-\tau) \tag{44}$$

i.e. the $f$ in the two half-cycles is the same. It follows from (43) that the $z(t)$ in the two half-cycles are in some sense a mirror of each other, through the dependency (43). Given (44), here we speculate that the physical cosmology starts and finishes in each half cycle, and that the second half cycle is a re-parsing of the first. By this we mean that not only is the scale factor of the Friedmann equation repeated (indirectly, by virtue of (22) and (43)), but also that the repetition extends down to the details of the matter and radiation, even as they depart from the perfect fluid approximation underlying the scale factor. In short, there is just one cosmological cycle, of duration $\tau$. This we call the 'no-copy' assumption. The alternative is that each half cycle is different in detail, even though it is an image of the preceding half-cycle at the level of the scale factor. Though we find this alternative unattractive, it cannot be ruled out.

The above conjecture shares with the Penrose view [1] that the Big Bang and the conformal singularity are closely related. But it does so in the context of a closed cosmology, rather than a repeating cycle. Indeed we argue given here that there is neither a meaningful physical distinction between each cycle, nor between two half-cycles.

## Dirac Wavefunction at the Conformal Boundary

It remains of course to justify this conjecture at the level of (non-homogeneous) matter and radiation fields. Here we offer an additional speculation involving the development of the Dirac field which seems to fit in with the no copy assumption.

It has been shown [4], that the only two components of the one-particle Dirac wavefunction survive to the conformal singularity. Prior to the boundary, the negative energy components are extinguished, and only positive components survive. The same analysis performed in the post-singularity era but reversed in time gives that only the negative energy components survive (moving towards the boundary from above). That is, in the Dirac representation

$$\psi(t,\mathbf{x}) = \begin{pmatrix} \phi(t,\mathbf{x}) \\ \chi(t,\mathbf{x}) \end{pmatrix} \tag{45}$$

then

$$\lim_{t\to t_{cs-}} \psi(t,\mathbf{x}) = \begin{pmatrix} \phi(t,\mathbf{x}) \\ 0 \end{pmatrix}, \quad \lim_{t\to t_{cs+}} \psi(t,\mathbf{x}) = \begin{pmatrix} 0 \\ \chi(t,\mathbf{x}) \end{pmatrix} \tag{46}$$

To obtain these limits, the bi-spinor was specified at a boundary condition some time away from the conformal singularity, and contained arbitrary contributions from both spinor components. If instead the boundary condition were situated at the conformal singularity, then an alternative interpretation of the above is that both spinors propagate away from the singularity in opposite directions in time. The $\phi(t,\mathbf{x})$ spinor propagates backwards in time and the $\chi(t,\mathbf{x})$ spinor propagates forwards. Inverting (46), only two components need be supplied very close to, and either side of, the boundary. Though their development away from the boundary may depend on the details, it will always be the case that there are only two effective degrees of freedom in the bi-spinor. Bearing in mind that (46) applies irrespective of the momentum there is no justification for choosing any particular momentum for the particle as it develops away from the boundary, so one may as well choose a wavefunction remains stationary in the Hubble frame. Then there will be no subsequent mixing of the two spinors – an initially zero spinor in the bi-spinor will remain zero thereafter (ignoring interactions).

## Mobius Band Conjecture

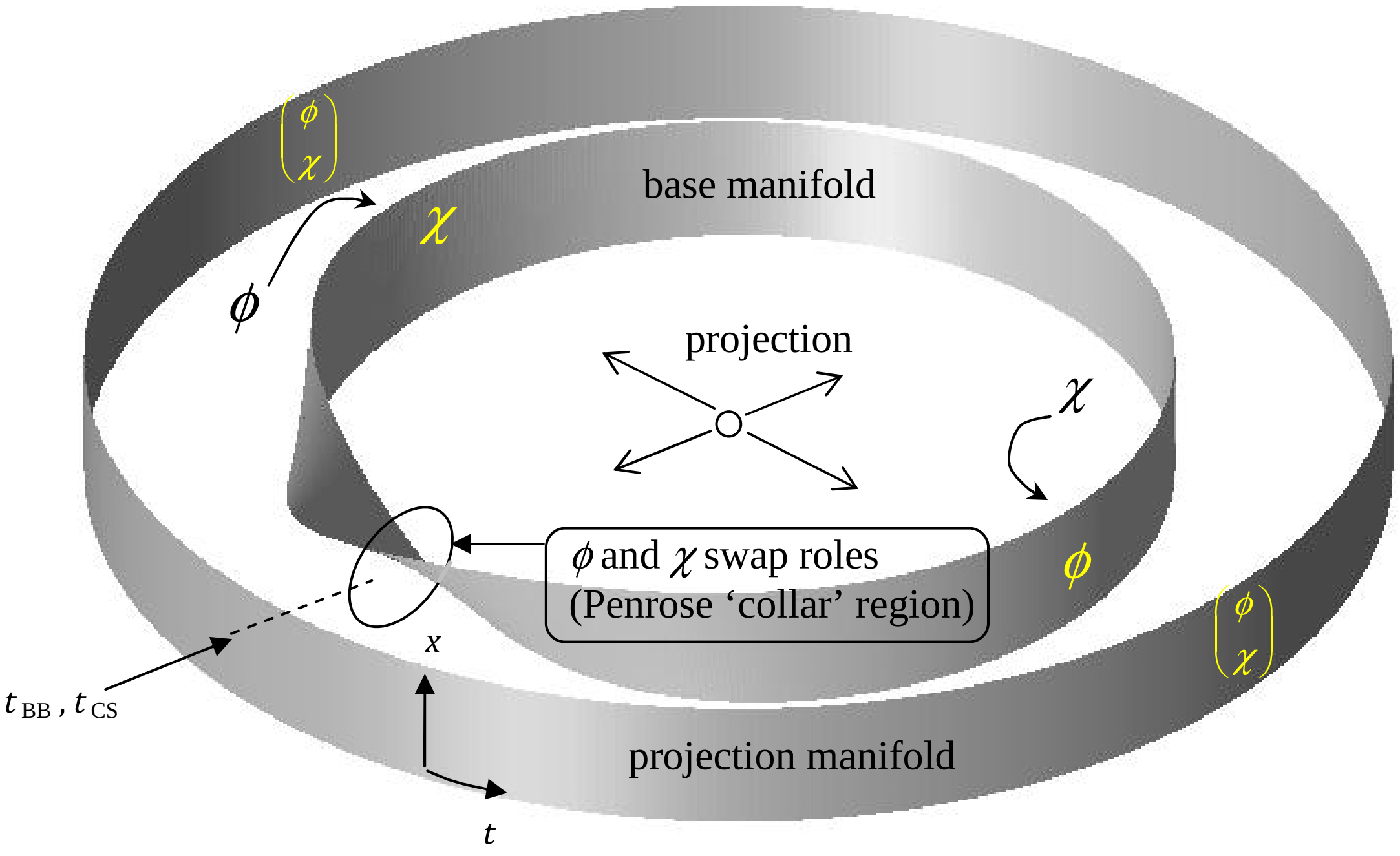

**FIGURE 1.** Manifold with Mobius topology. The bi-spinor wavefunction is presumed to be a local projection that combines spinor fields from both sides of the manifold at a single spacetime location in the traditional (single-sided, orientable) manifold. The two spinor fields exchange roles within the wavefunction within a 2π trip around the band corresponding to the time interval between the Big Bang and the conformal singularity. They return to their starting configuration after 4π.

Recalling the assertion above that the two half cycles are physically the same, the behavior of the Dirac function near the boundary suggests the arrangement depicted in Fig. 1. The two-sided base manifold (actually its projection onto a single-sided cylinder as show in the figure) has the advantage of explaining the duplication of the two half-cycles that are intrinsic to the Friedmann equation, yet remaining consistent with the 'no copy' assumption, i.e. without doubling the volume of spacetime required to describe it, and without, therefore the danger of a physically redundant mirror-image cosmology. The twist in the manifold ensures that however the scale factor is assigned to the two-sided manifold, it will have a period exactly twice that of the conformal duration, as required. It is assumed that the conventional spacetime of cosmology comprises the projection manifold and the standard cosmological metric even though the implication is that the Mobius manifold is more fundamental.

# REPEATED ROOTS

## Relationship to Self-Similarity

The above self-similarity conjecture does not by itself impose a constraint on the Friedmann equation because it permits on a Mobius transformation to relate the self-similar form to the traditional form (6). Here we investigate the possibility of a greater symmetry that imposes constraints on the traditional form (6), that the roots of the quartic in $z(t)$ occur in repeated pairs. If so, then the right hand side of (27) is a perfect square. In that case we will see that the closed form solution for the scale factor will be expressible in terms of log or inverse trigonometric or inverse hyperbolic functions, rather than Elliptic functions. Since this is a higher symmetry, we expect tighter constraints on the coefficients $\Gamma_i$, and possibly constraints on the $\Omega_i$. Note however that the preceding arguments, including the alleged role self-similarity in the Friedmann equation and the likely existence therefore of an $\Omega_3$ species are independent of the following speculation.

## Template

Consider the possibility that (6) is expressible as

$$\frac{1}{H^2}\left(\frac{da(t)}{dt}\right)^2 = \kappa^2\left(\left(a(t)-a_0\right)^2+b\right)^2 \tag{47}$$

I.E. that each root of the quartic on the right hand side of (6) is repeated twice. Note there are three arbitrary constants $\kappa, a_0, b$ on the right hand side of (47).

## Relationship to the Mobius Transformation

It can be shown that under the presumed form (47) the Mobius transformation (22) does not introduce any new degrees of freedom, but rather that $a_0, \kappa, \lambda$ become functions of $\alpha,\gamma,\delta$. One way to see this is to note that (47) can be integrated to give

$$\kappa H\left(t-t_0\right) = \frac{1}{\sqrt{b}}\tan^{-1}\left(\frac{a-a_0}{\sqrt{b}}\right) \tag{48}$$

and therefore

$$\begin{aligned} a(t) &= a_0 + \sqrt{b}\tan\left(\sqrt{b}\kappa H(t-t_0)\right) \\ &= a_0 + \sqrt{b}\frac{\chi - z(t)}{1+\chi z(t)}; \quad z(t) := \tan\left(\sqrt{b}\kappa Ht\right), \quad \chi := \tan\left(\sqrt{b}\kappa Ht_0\right) \\ &= \frac{a_0 + \sqrt{b}\chi + \left(\chi - \sqrt{b}\right)z(t)}{1+\chi z(t)} \end{aligned} \tag{49}$$

which has the form (22) (after dividing numerator and denominator by $\chi - \sqrt{b}$ ). Since two concatenated Mobius transformation is another Mobius transformation, (22) is already contained in the form (47). Hence there is no further benefit to allowing a Mobius transformation if the template form (47) is enforced. Therefore (47) can be applied directly to (6) rather than (27).

This finding suggests that the template (47) could be motivated by an a priori assumption that the Friedmann equation express precisely invariance under Mobius transformations (no more and no less).

## Inferred Dependencies

Applying the template (47) to (6) one has immediately that

$$\kappa^2 = \Omega_v, \quad \kappa^2\left(b + a_0^{\ 2}\right)^2 = \Omega_r \tag{50}$$

Continuing with the expansion, and applying (50), one finds

$$\Omega_c = \frac{\Omega_m^2}{4\Omega_r} + W_\sigma, \quad \Omega_3 = \frac{\Omega_m}{\Omega_r}W_\sigma, \quad \Omega_v = \frac{W_\sigma^{\ 2}}{4\Omega_r} \tag{51}$$

where

$$W_\sigma := 2\sigma\sqrt{\Omega_r} - 2\Omega_r - \Omega_m, \quad \sigma = \pm 1 \tag{52}$$

and we have chosen to express the other contributions in terms of the radiation and matter contributions. The deceleration (11) is found to be

$$q_0 = \sigma\sqrt{\Omega_r}\left(1 - \frac{W_\sigma}{2\Omega_r}\right) = 2\Omega_r + \Omega_m - \frac{W_\sigma}{2\Omega_r}\left(W_\sigma + \Omega_m\right) \tag{53}$$

## Predictions in the Case of Zero Curvature

We examine at first the consequences of an a priori assumption that the curvature is zero. From (51)

$$\frac{\Omega_m^2}{4\Omega_r} + W_\sigma = \frac{\Omega_m^2}{4\Omega_r} + 2\sigma\sqrt{\Omega_r} - 2\Omega_r - \Omega_m = 0 \Rightarrow \Omega_m = 2\Omega_r\left(1 \pm \sqrt{3 - 2\sigma / \sqrt{\Omega_r}}\right) \qquad (54)$$

For $\Omega_r \sim 10^{-4}$, $\Omega_m$ is real only for $\sigma = -1$, and positive only for the positive root in (54). Hence

$$\Omega_m = 2\Omega_r\left(1 + \sqrt{3 + 2 / \sqrt{\Omega_r}}\right) \qquad (55)$$

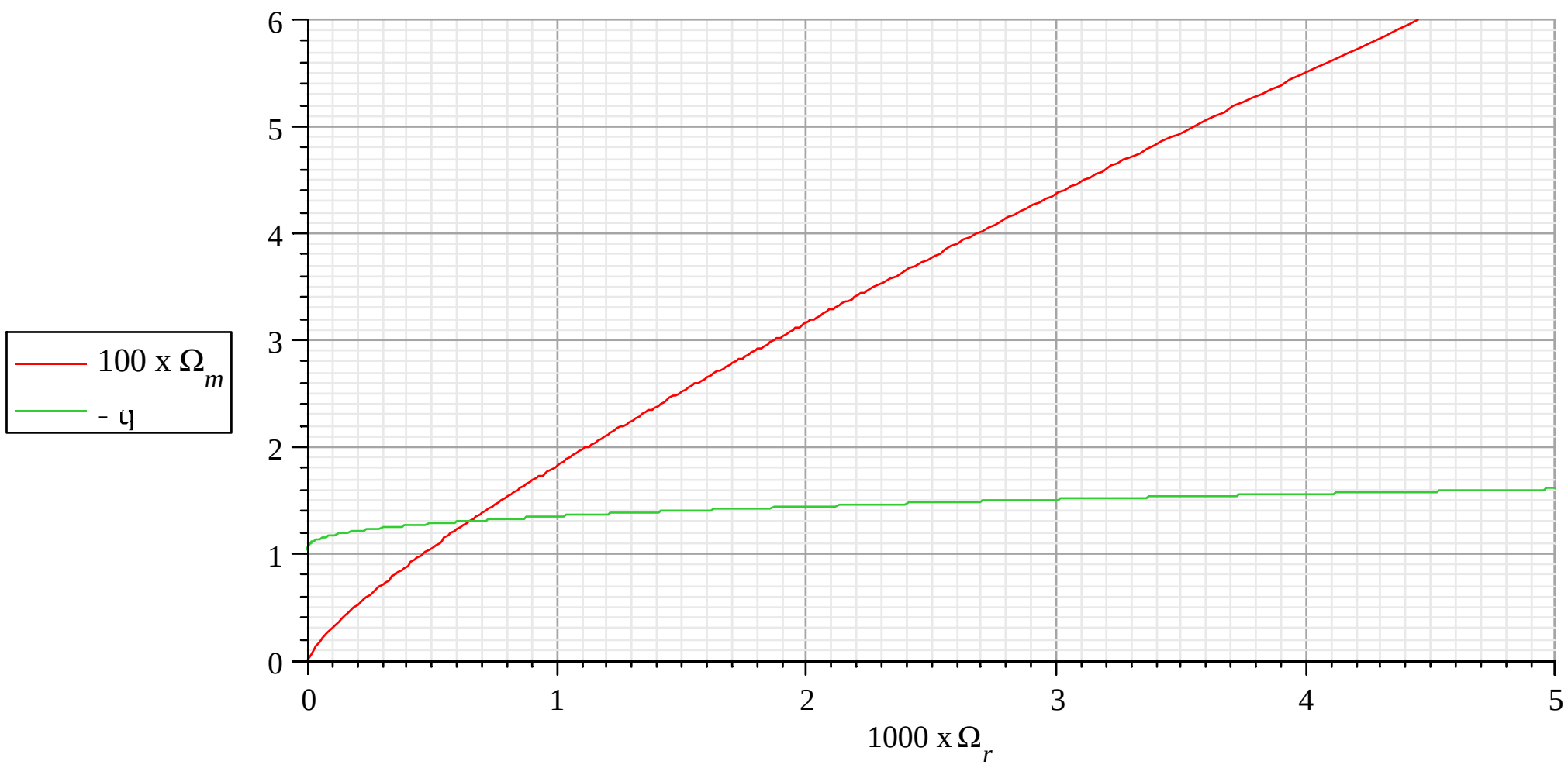

**FIGURE 2.** $\Omega_m$, $q_0$ for various $\Omega_r$.

It is clear from the figure that there is no room for any substantial amount of cold dark matter. Further, the radiation density would have to be more than an order of magnitude above the book-value of the EM density. Specifically, at $\Omega_m = \Omega_b = 0.04$ we would require $\Omega_r = 0.0027$. This implies a significant role for relativistic neutrinos, or an additional species of radiation. At these values, (51) gives

$$\Omega_r = 0.0027, \quad \Omega_m \approx \Omega_b = 0.04, \quad \Omega_c = 0, \quad \Omega_3 = -1.1, \quad \Omega_v = 2.1, \quad q_0 = -1.5 \qquad (56)$$

If instead we can accept $\Omega_m = 0.02$, then

$$\Omega_r = 0.0011, \quad \Omega_m \approx \Omega_b = 0.02, \quad \Omega_c = 0, \quad \Omega_3 = -0.79, \quad \Omega_v = 1.77, \quad q_0 = -1.25 \qquad (57)$$

Notice that $\Omega_3$ is negative. What are the implications for structure formation in this case?

## *Curvature*

The mass-radiation relation for various curvatures is given in Figures 3, 4, and 5. Figure 3 and 4 are different regions of the parameter space for the same roots selection $\sigma = 1$. Notice how in Figure 3 how $\Omega_c = 0.3$ generates of which is closed, near the origin. In fact $\Omega_c = 0.1$ generates a still tighter closed curve which is not visible at this scale. Figure 4 expands the region near the origin showing how more modest curvatures are constrained to more modest values of $\Omega_r$. Negative values of mass are included to aid visualization of the relationships without intending to imply that the net mass-energy-density is negative.

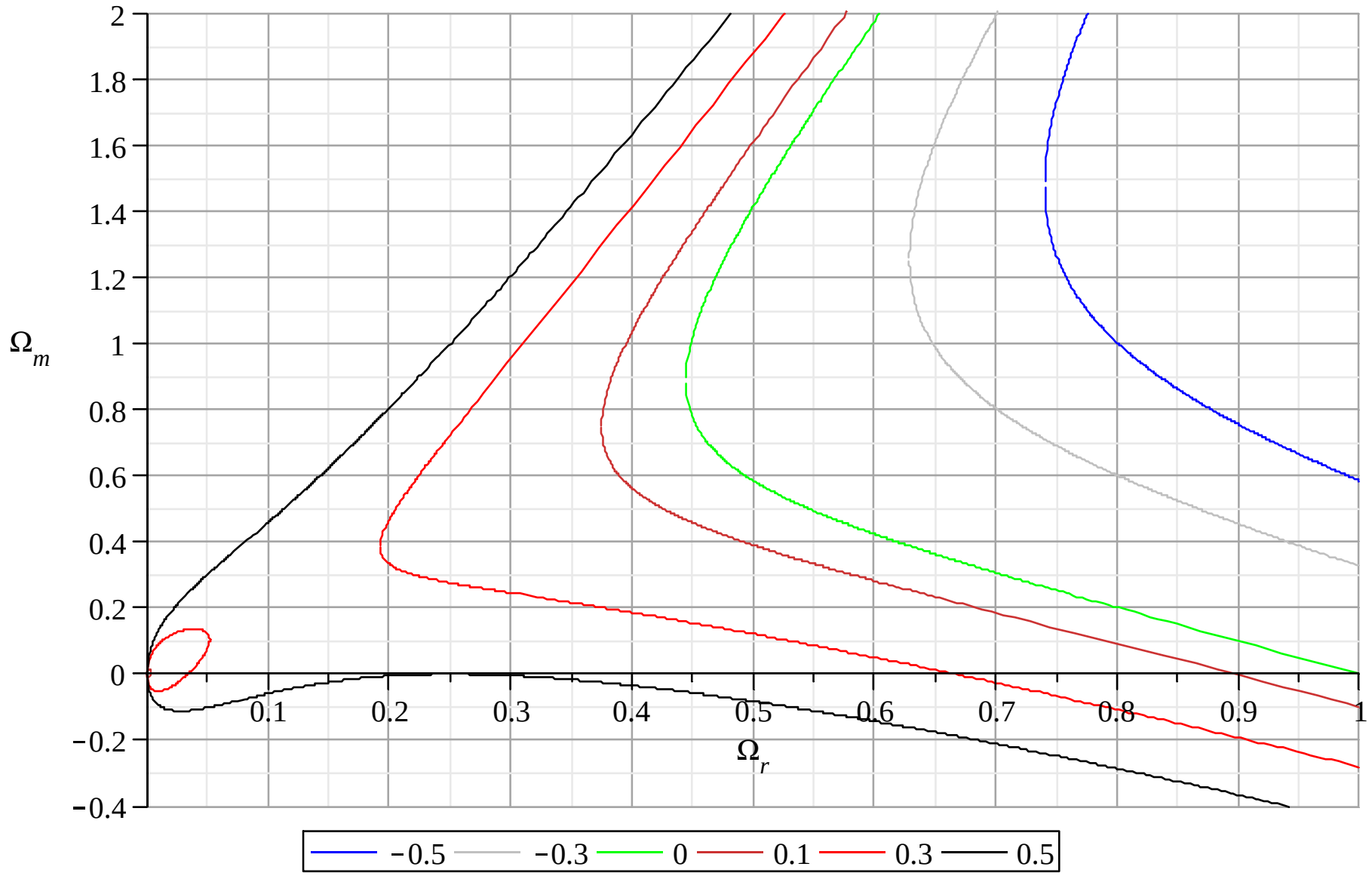

**FIGURE 3.** $\Omega_m$, $\Omega_r$ relation for various $\Omega_c$ at large $\Omega_r$; $\sigma = 1$.

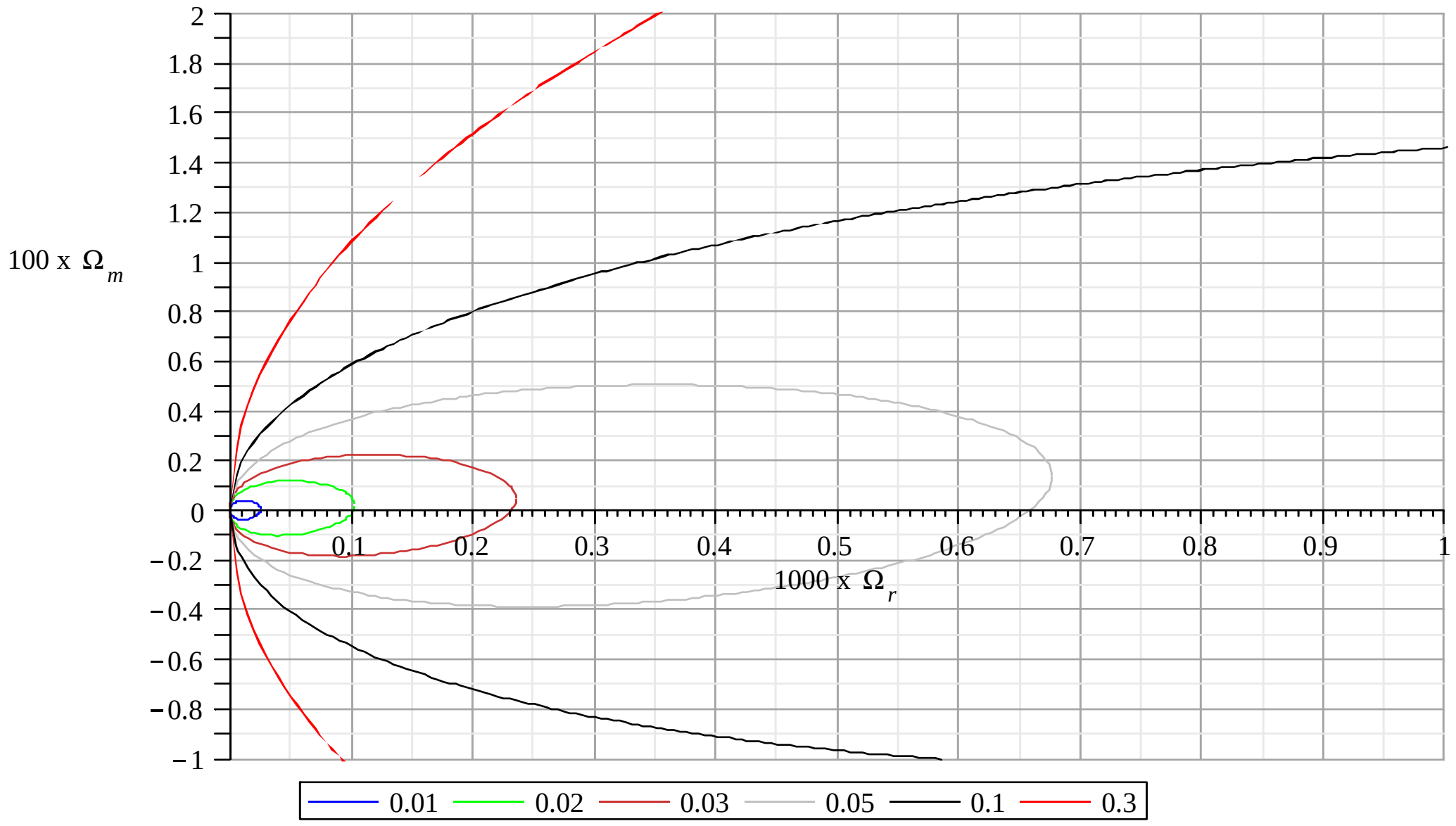

**FIGURE 4.** $\Omega_m$, $\Omega_r$ relation for various $\Omega_c$ at modest $\Omega_r$; $\sigma = 1$.

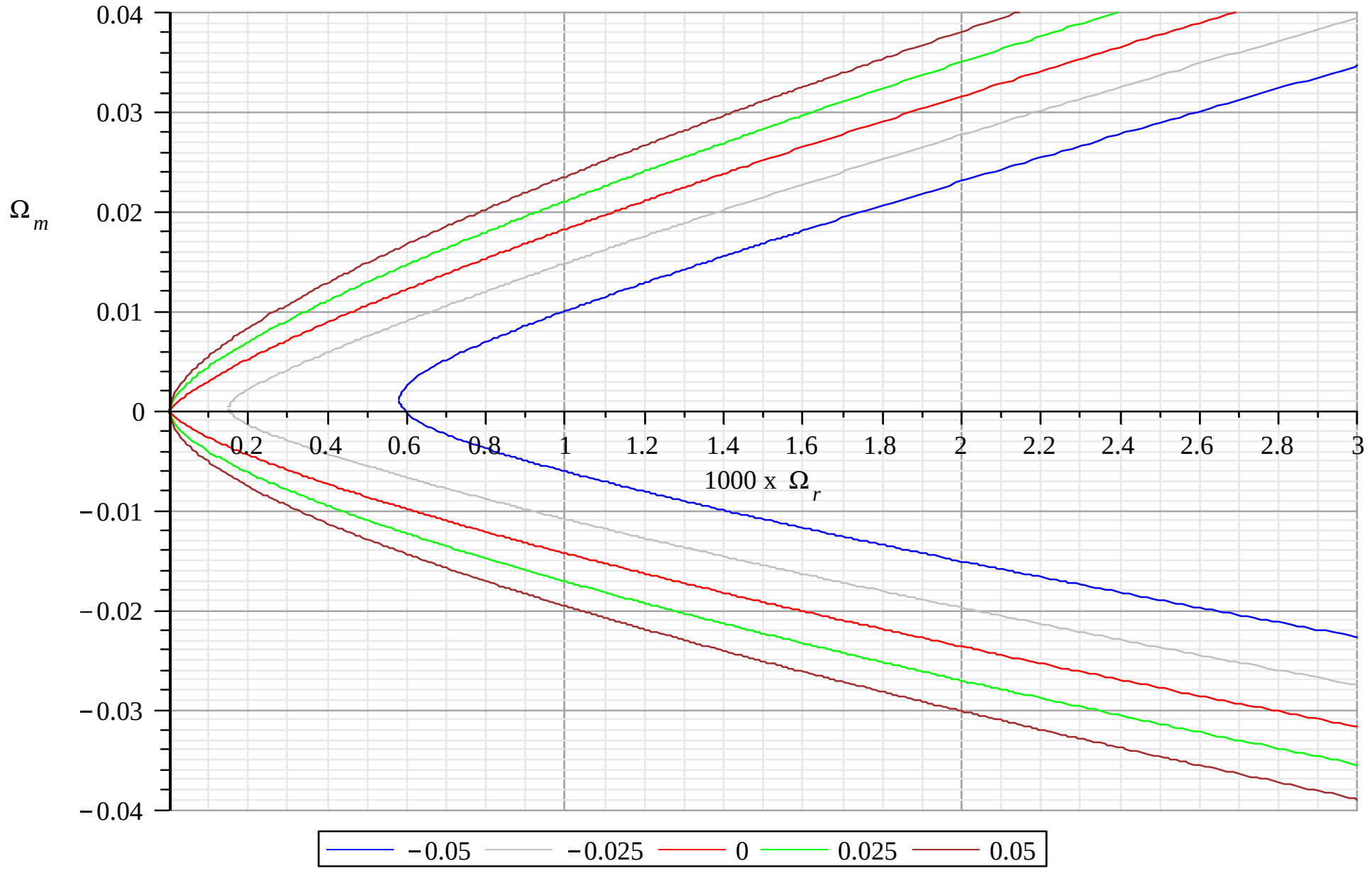

**FIGURE 5.** $\Omega_m$, $\Omega_r$ relation for various $\Omega_c$ at modest $\Omega_r$; $\sigma$ = -1.

It is clear that only $\sigma$ = -1 is tolerable, so only that case will be considered henceforth.

## $\Omega_3$ *in the General Case*

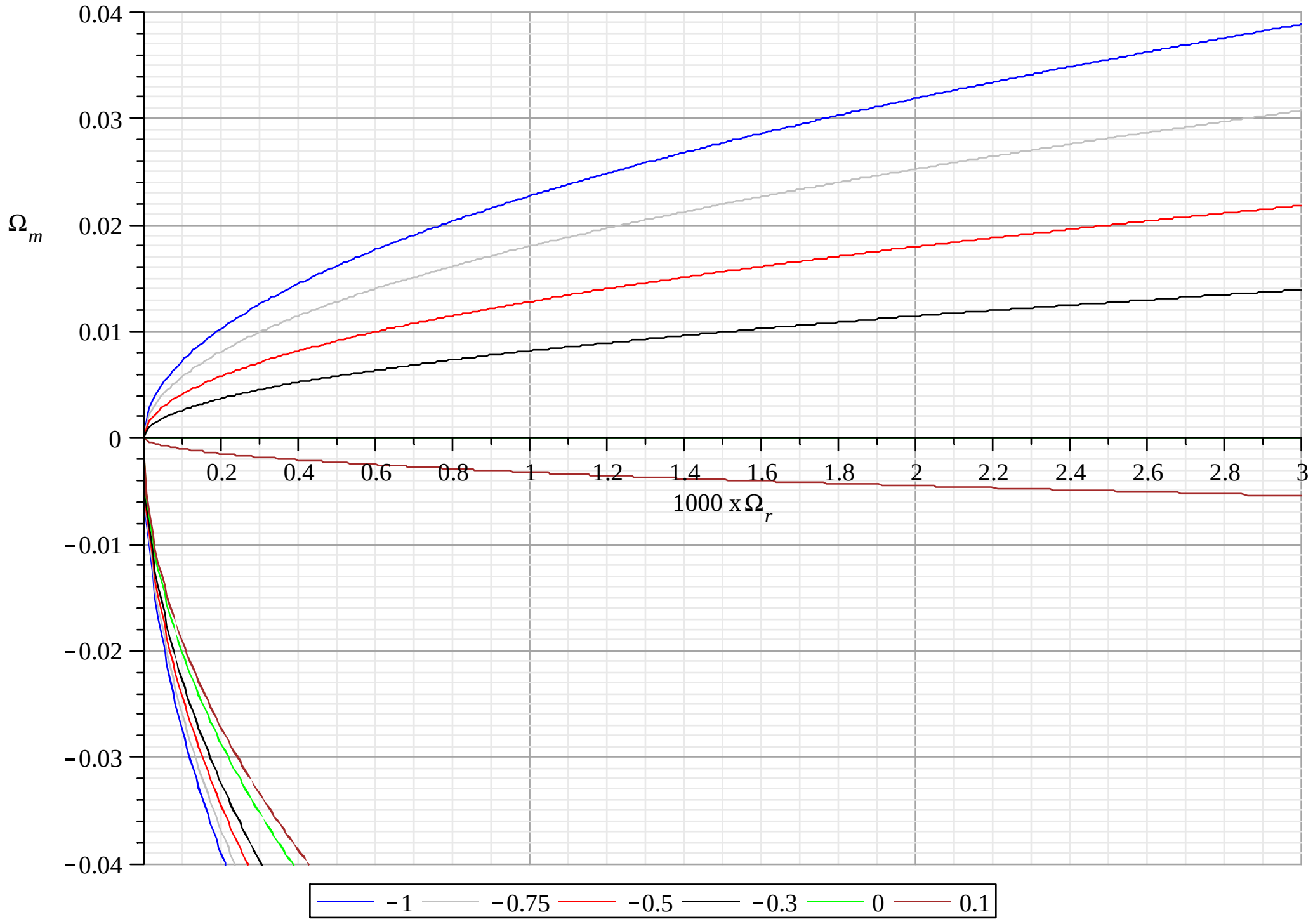

**FIGURE 6.** $\Omega_m$, $\Omega_r$ relation for various $\Omega_3$; $\sigma$ = -1.

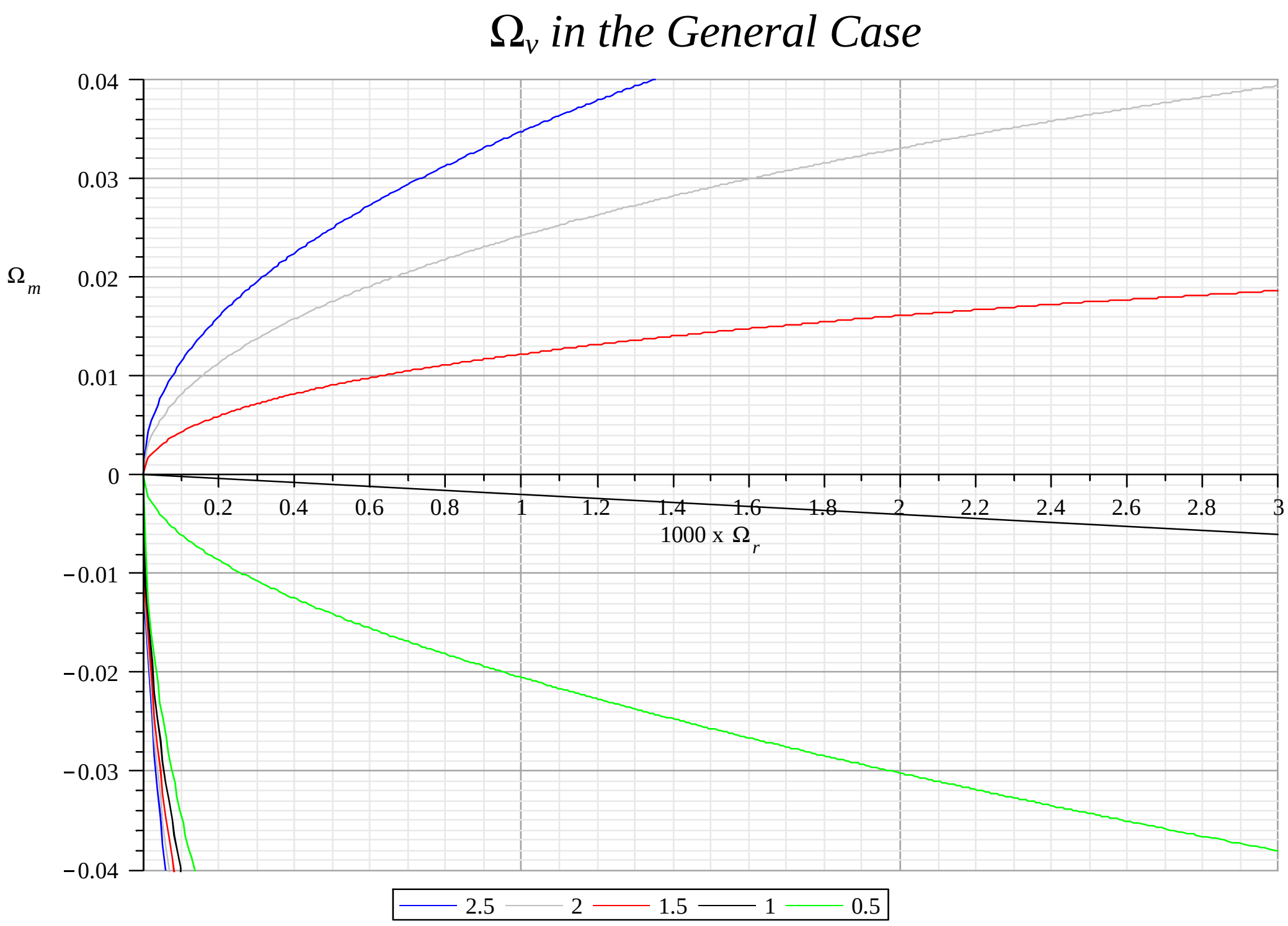

**FIGURE 7.** $\Omega_m$, $\Omega_r$ relation for various $\Omega_v$; σ = -1.

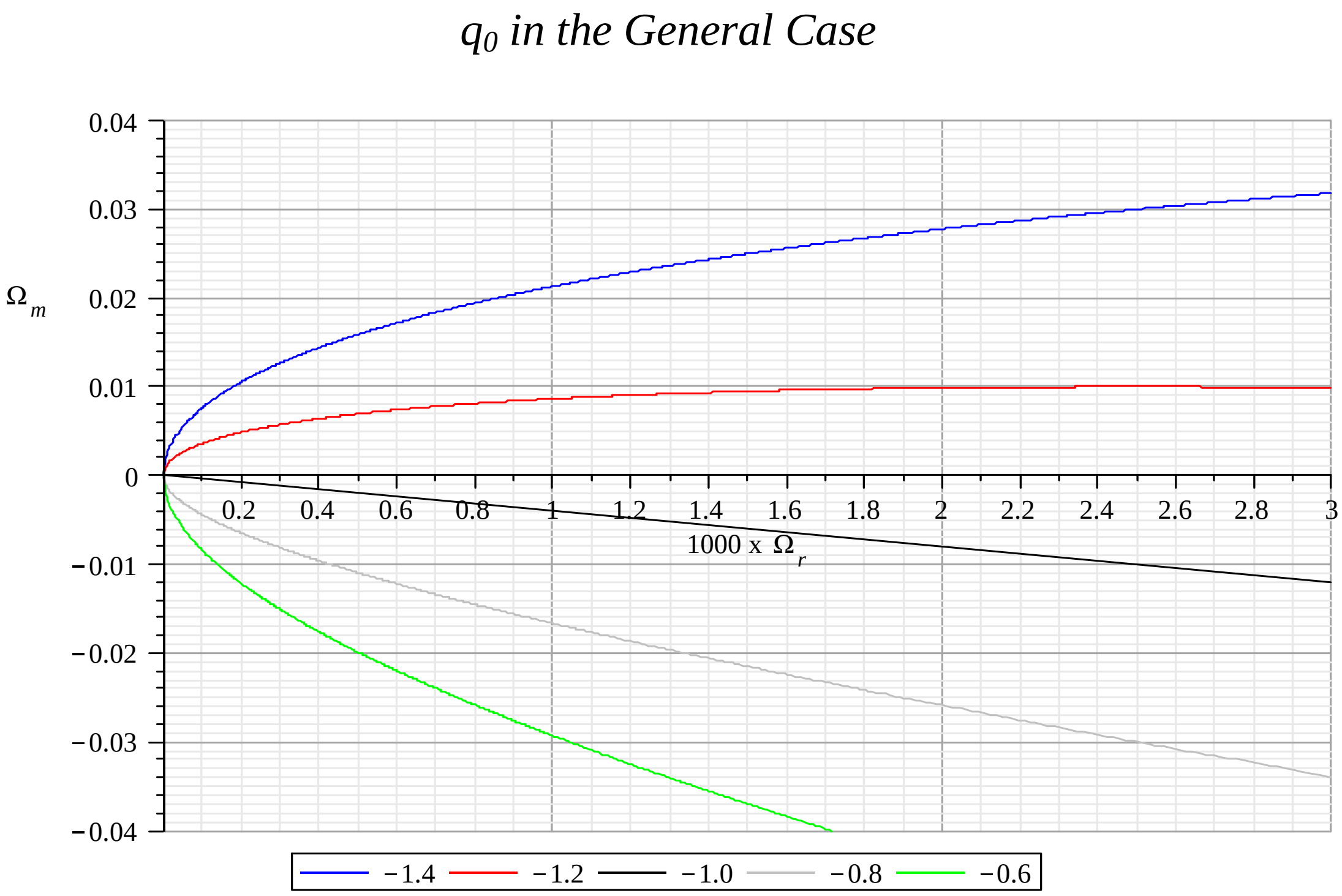

**FIGURE 8.** $\Omega_m$, $\Omega_r$ relation for various $q_0$; σ = -1.

Solving (53) for $\Omega_m$ and setting $\sigma$ = -1 one obtains

$$\Omega_m = -2(1+q_0)\sqrt{\Omega_r} - 4\Omega_r \tag{58}$$

This relation is shown in Fig. 8.

### *Conclusions on the Repeated Roots*

The only conceivably acceptable set of parameters is centered on $\Omega_m$ close to the observed value with no CDM. The radiation is much higher than the standard value, though this may not be a problem if due to very light neutrinos. For $q_0$ close to -1 the model fit requires that the vacuum energy is closer to 2 than 1, and the conjectured $\Omega_3$ term has a negative energy around -0.75. For the model fit to work it seems that this term must play the role in structure formation normally attributed to CDM.

## SUMMARY

The Friedmann equation is form invariant under $a(t) \to \pm 1/a(t)$ and more generally under a Mobius transformation of the scale factor. Defining a synthetic quantity $z(t)$

$$z(t) = -\frac{\alpha - \gamma a(t)}{\beta - \delta a(t)} \tag{59}$$

a modified Friedmann equation in $z(t)$ can be made invariant under $z(t) \to \pm 1/z(t)$ by suitable choice of constants $\alpha, \beta, \gamma, \delta$. We have looked into the possibility that this might be the more fundamental equation, from which the traditional Friedmann equation in $a(t)$ is obtained via the inversion of (59). If so, it strongly suggests that the traditional Friedmann equation is missing a species with equation of state $p = -2\rho/3$.

It was shown that the scale factor is periodic. And it was shown that the scale factors in adjacent half-cycles are simply related, especially if expressed in terms of $z(t)$. It was suggested that one half-cycle represents a full cosmology, the second being a re-parsing of the first. A sketch was made of the evolution of a Dirac wavefunction on a Mobius band having some qualities consistent with this point of view.

We considered also the much tighter constraint that the right hand side of the Friedmann equation is a perfect square. The cosmology it predicts is interesting, but far from the standard model, most notably that there is no room for dark matter.

# ACKNOWLEDGEMENTS

The author is indebted to Anthony Lasenby and J. Alberto Vazquez at the Kavli Institute for Cosmology for donating their time and expertise, especially in looking for evidence of the $\Omega_m$ species.

# APPENDIX A
# TRANSFORMATION OF RW CURVED SPACETIME TO CONFORMAL COORDINATES

To convert (2) to the conformal form it is useful first to make the transformation

$$U = \tan u, \quad V = \tan v \tag{A1}$$

where u,v and U,V are light-cone coordinates:

$$u = \frac{k}{2}(t - t_0 + r), \quad v = \frac{k}{2}(t - t_0 - r), \quad U = \frac{K}{2}(T - T_0 + R), \quad V = \frac{K}{2}(T - T_0 - R). \tag{A2}$$

Note the (new) time offsets that are freedoms of the transformation. Henceforth we will take these as implicit and suppress $t_0$ and $T_0$. Substituting (A2) and (A1) into (2) gives

$$ds^2 = A^2(T,R)\left(dT^2 - dR^2 - R^2 d\Omega^2\right) \tag{A3}$$

where

$$A(T,R) = \frac{K}{k} a(t) \cos u \cos v. \tag{A4}$$

It remains to express the right hand side in terms of the new coordinates. Eqs.(A1) give

$$\cos u = \frac{1}{\sqrt{1+U^2}}, \quad \cos v = \frac{1}{\sqrt{1+V^2}} \tag{A5}$$

so that

$$\cos u \cos v = \frac{1}{\sqrt{\left(1+\left(\frac{K}{2}(T+R)\right)^2\right)\left(1+\left(\frac{K}{2}(T-R)\right)^2\right)}} = \frac{1}{\sqrt{\left(1-K^2X^2/4\right)^2+K^2T^2}} \tag{A6}$$

where $X^2 = T^2 - R^2$. The scale factor argument is converted using

$$kt = u + v = \tan^{-1} U + \tan^{-1} V = \tan^{-1}\left(\frac{K}{2}(T+R)\right) + \tan^{-1}\left(\frac{K}{2}(T-R)\right). \tag{A7}$$

The two inverse tangents can be combined as follows:

$$\tan(kt) = \frac{KT}{1-\frac{K^2}{4}\left(T^2-R^2\right)} \Rightarrow kt = \tan^{-1}\left(\frac{KT}{1-K^2X^2/4}\right). \tag{A8}$$

Using (A7) and (A6) in (A4) gives the relationship between the two scale factors

$$A(T,R) = \frac{K}{k}\frac{a\left(\frac{1}{k}\tan^{-1}\left(\frac{KT}{1-K^2X^2/4}\right)\right)}{\sqrt{\left(1-K^2X^2/4\right)^2+K^2T^2}} \tag{A9}$$

Reversion to flat space is achieved by letting

$$k = K; \quad K \to 0 \Rightarrow A(T,R) \to a(T). \tag{A10}$$